\documentclass[twocolumn,superscriptaddress,floatfix,preprintnumbers,nofootinbib,prd,10pt]{revtex4-2}
\usepackage{docs}%
\usepackage{bm}%
\usepackage[colorlinks=true,citecolor=teal]{hyperref}
\pdfoutput=1
\usepackage{graphicx}
\usepackage{xcolor,colortbl}
\usepackage{subfig}
\usepackage{dcolumn}
\usepackage{tikz}
\usepackage{physics}
\usepackage{mathtools}
\usepackage{bm}
\usepackage{amssymb}
\usepackage{amsmath}
\usepackage{epsfig}
\usepackage[toc,page]{appendix}
\usepackage{slashed}
\usepackage{hhline}
\usepackage{color}
\usepackage{amsmath} 
\usepackage{upgreek} 
\usepackage{bm} 
 \usepackage{multirow}
\usepackage{url}
\usepackage{empheq}
\def\be{\begin{equation}}
\def\ee{\end{equation}}
\def\bea{\begin{eqnarray}}
\def\eea{\end{eqnarray}}
\def\ba{\begin{array}}
\def\ea{\end{array}}
\def\ben{\begin{enumerate}}
\def\een{\end{enumerate}}

\def\bw{\begin{widetext}}
\def\ew{\end{widetext}}

\newcommand{\dsl}{\pa \kern-0.5em /} 

\def\b{\beta}

\def\pa {\partial}

\def\nn{\nonumber}

\begin{document}
\allowdisplaybreaks

\title{Entanglement wedge method, out-of-time-ordered correlators, and pole-skipping}

\author{Banashree Baishya}
\email[E-mail: ]{b.banashree@iitg.ac.in}
\affiliation{Department of Physics, Indian Institute of Technology Guwahati, Assam-781039, India}

\author{Adrita Chakraborty}
\email[E-mail: ]{achakraborty@agh.edu.pl}
\affiliation{Faculty of Physics and Applied Computer Science, AGH University of Krakow, al. A. Mickiewicza 30, 30-059 Krakow, Poland} 

\author{Nibedita Padhi} 
\email[E-mail: ]{nibedita.phy@iitkgp.ac.in}
\affiliation{Department of Physics, Indian Institute of Technology Kharagpur, Kharagpur,721302, India} 

\begin{abstract}
   We investigate two salient chaotic features, namely Lyapunov exponent and butterfly velocity, for an asymptotically Lifshitz black hole background with arbitrary dynamical critical exponent. These features are computed using three methods: entanglement wedge method, out-of-time-ordered correlator computation and pole-skipping. We present a comparative study where all of these methods yield exactly similar results for the butterfly velocity and Lyapunov exponent. This establishes an equivalence between all three methods for probing chaos in the chosen gravity background. Furthermore, we evaluate the chaos at the classical level by computing the eikonal phase and Lyapunov exponent from the bulk gravity. In the classical approach, we comment on potential limitations while choosing the turning point of the null geodesic in our gravity background. All chaotic properties emerge as nontrivial functions of the anisotropy index. By examining the classical eikonal phase, we uncover different scattering scenarios in the near-horizon and near-boundary regimes. Finally, we remark on a possible classical/quantum correspondence from the analysis of classical eikonal phase shift and Lyapunov exponent.
\end{abstract}


\maketitle
\section{Introduction}
The presence of chaos is robust in various disordered many-body systems, as evidenced by the study of black holes. As an immediate consequence of chaos, there occurs dynamical growth of local perturbations in the disordered systems due to sufficiently small changes in their initial conditions. This results in considerable deviations in the trajectory of any initial points. The Anti-de Sitter (AdS)/Conformal Field Theory (CFT) correspondence \cite{Maldacena:1997re, Aharony:1999ti} provides one of the most competent platforms to realise quantum gravity through a conformally invariant strongly interacting quantum field theory that lives on the boundary of the dual AdS background. Such holographic duality can also be extended as a general gauge/gravity duality where the field theory breaks conformal invariance and its dual geometry is eventually a non-AdS one. A class of nonrelativistic Lifshitz field theories are a theory that follows Lifshitz scaling symmetry
\begin{equation}
     t\rightarrow \Omega^\xi t,~~\vec{x}\rightarrow \Omega\vec{x},
\end{equation}. lifshitz scaling is generally characterized by the dynamical critical exponent $\xi\in \mathbb{R}$, $\xi\geq 1$. Such QFTs break Lorentz invariance due to the presence of an anisotropic scale factor along the temporal direction. Hence, they do not adhere to CFT. The holographic dual of Lifshitz field theories is the nonrelativistic Lifshitz geometry, which shares similar scale symmetry \cite{Bala, Walter, Juan08, Allan08, Kachru}. Exploring such theories finds its significance in the context of the correlation between high energy physics and condensed matter physics, as these theories, especially for $\xi=2$, are directly related to strongly correlated electron systems. A series of research for a rigorous understanding of Lifshitz field theories has been going on over the last decade by implementing various perspectives \cite{Ker2017, KORD, parkChanyong, Cheyne:2017bis, Basak:2023otu}. Theories with Lifshitz scaling exhibit chaos due to the presence of finite anisotropy. Previously, the probe circular strings moving in the holographic Lifshitz bulk spacetime have been found to possess chaotic behaviour\cite{Bai:2014wpa}. In addition to this, the nonintegrability of the class of hyperscale violating Lifshitz theories is studied in \cite{Giataganas:2014hma}. Apart from classical chaos, the chaotic growth of marginally irrelevant operators in the 2+1 D quantum Lifshitz model with $\xi=2$ is previously explored in \cite{Plamadeala_2018} by using a perturbative approach. \cite{Sircar:2016old} presents extensive Holographic studies on scrambling properties for a general class of gravitational theories, including the asymptotically Lifshitz black hole. The latter reads as the plausible gravity dual of the finite temperature counterpart of Lifshitz field theories\cite{Bala}.  

There are many ways to analyse quantum chaos, such as Random matrices \cite{Cotler:2016fpe}, Complexity \cite{Susskind:2014rva, Brown:2015lvg, Magan:2018nmu}, Ergodicity \cite{Pausch_2021, ergo}, etc. In a recent study\cite{Dong:2022ucb}, the butterfly effects of higher derivative gravity theories are investigated by two equivalent holographic understandings. One is using the method of entanglement wedge reconstruction, and the other one is to incorporate localized gravitational shock waves at the horizon of the corresponding dual black hole. In this article, we aim to study the chaotic properties of such a theory from the bulk perspective. We will implement three well-known holography-based methods of computing quantum chaos in bulk gravity and check the equivalence among these for the asymptotically Lifshitz black hole. Namely, these methods are:
\begin{itemize}
    \item Entanglement wedge method \cite{Mezei:2019zyt, Fischler:2018kwt, Dong:2022ucb} 
    \item Out of time-ordered correlation function (OTOC) \cite{Shenker:2013pqa, Maldacena:2015waa, Shenker:2013yza, Roberts:2014isa, Shenker:2014cwa, Plamadeala_2018, Roberts:2014ifa, Khetrapal:2022dzy, Lin_2018, Gu:2016hoy, Das:2022jrr}
    \item Pole-skipping \cite{Blake:2017ris, Blake:2018leo, Grozdanov:2018kkt, Ahn:2019rnq, Blake:2021hjj, Sil:2020jhr, Li:2019bgc, Amano:2022mlu, Abbasi:2019rhy, Baishya:2023mgz, Baishya:2023ojl, Yuan:2020fvv}
\end{itemize}
\begin{enumerate} 
    \item In AdS/CFT correspondence, entanglement wedge reconstruction \cite{entang, Wall:2012uf, Czech:2012bh} depicts that all the information of the entanglement wedge in the bulk lies in the boundary region. When we perturb the boundary state by a local operator and let the system evolve, the information gets scrambled in the whole space at a late time. This information propagates outward with a constant velocity. The whole scenario has a bulk description. Perturbing the boundary means probing a particle that falls toward the black hole in the bulk close to the asymptotic boundary. The trajectory of the infalling particle then reaches inside the extremal surface, also known as the Ryu-Takayanagi(RT) surface \cite{rt}. As the trajectory of the particle changes, the RT surface changes its shape. At a late time, the RT surface reaches the near-horizon region of the black hole with a constant velocity called butterfly velocity ($v_B$). This method is very well-known for extracting the butterfly velocity for any disordered holographic system \cite{Mezei:2016wfz}. This method is further understood in higher-derivative gravity theories\cite{Dong:2022ucb}. In our work, we employ this method to calculate the butterfly velocity in the Einstein-Proca-type Lifshitz gravity theory. The action of such theories contains a massive vector field that breaks Lorentz symmetry\cite{Taylor:2015glc}. 
    
 \item Quantum chaos can further be studied by calculating the double commutator of any two generic operators $V$ and $W$ as,

    \begin{align}\label{commmu}
        & C(t)=\langle V^{\dagger}(0)W^{\dagger}(t)W(t)V(0)+V^{\dagger}(0)W^{\dagger}(t)W(t)V(0)\nonumber\\&\quad\quad\quad -\underbrace{W^{\dagger}(t)V^{\dagger}(0)W(t)V(0)-V^{\dagger}(0)W^{\dagger}(t)V(0)W(t)\rangle_{\beta}}_{\text{out-of time-ordered}}\,.
    \end{align}
For a chaotic system, the above commutator takes an exponential form as \cite{Shenker:2013pqa},
\begin{equation}
    C(t)\approx\exp{\lambda_{L}t} \,,
\end{equation}
where the exponent $\lambda_L$ is the Lyapunov exponent and is always positive for a chaotic system. Study of the commutator in holographic systems for localized perturbations \cite{Shenker:2013pqa, Shenker:2014cwa} yields,
\begin{equation}
     C(t,x)\approx \exp{\lambda_{L}\left(t-t_{*}-\frac{|x|}{v_B}\right)}\,\label{commutator},
     \end{equation}
where the operators are separated by a spatial distance $x$. The exponent $v_B$ is the butterfly velocity and $t_*$ is the scrambling time, at which $C(t)$ becomes of order $\mathcal{O}(1)$. The out-of-time-ordered correlators (OTOC) show a large value until the "scrambling time" \cite{Sekino:2008he, Hayden:2007cs} and decrease rapidly thereafter. The decrease of OTOC leads to an increase of $C(t)$. This signifies the butterfly effect. OTOC has been rigorously studied in connection to black holes and quantum chaos \cite{Shenker:2013pqa, Maldacena:2015waa, Shenker:2013yza, Roberts:2014isa, Shenker:2014cwa}. To mention the field theory perspective, OTOC has been computed in 2D CFT \cite{Roberts:2014ifa, Khetrapal:2022dzy}, Ising chain \cite{Lin_2018}, rational CFT \cite{Gu:2016hoy}, driven CFT \cite{Das:2022jrr} and numerically in the quantum Lifshitz model \cite{Plamadeala_2018}. In two-derivative gravity, the Lyapunov exponent is bounded as \cite{Maldacena:2015waa, Mezei:2016wfz},
\begin{equation}\label{bound}
    \lambda_{L}\leq\frac{2\pi}{\beta}\,
   \end{equation}
where $\beta$ is the inverse temperature. As a natural question, one can ask whether a theory with broken Lorentz invariance can lead to a violation of this Lyapunov exponent \eqref{bound}. Again, the bound on the butterfly velocity is also a point of interest here. Many works \cite{Fischler:2018kwt, Giataganas:2017koz, Gursoy:2020kjd, Eccles:2021zum} have been done in these directions.
\item The chaotic nature we discussed in the previous two methods can be captured in the properties of a two-point energy density correlation function at the \textit{pole-skipping} \cite{Blake:2017ris, Blake:2018leo, Grozdanov:2018kkt, Ahn:2019rnq} point. Pole-skipping points are holographically related to the near-horizon properties of the metric perturbation in bulk. Pole-skipping (P-S) occurs when the lines of poles and zeros of retarded Green's function intersect, i.e., a would-be pole gets skipped! At the frequency and momentum specified at the P-S points, Green's function is not uniquely defined \cite{Blake:2019otz, Natsuume:2019xcy}. These values of the frequency ($\omega_*$) and momentum ($k_*$) in the energy-density two-point function are connected to the measure of chaos \cite{Blake:2017ris, Blake:2018leo} through the relations
\begin{equation}\label{poleskip}
\omega_{*}=i\lambda_{L},\hspace{2cm}k_{*}=\frac{i\lambda_{L}}{v_B} ,
\end{equation}
where $\lambda_{L}$  and $v_B$ are the Lyapunov exponent and the butterfly velocity respectively. In this paper, we have checked whether pole-skipping is a good diagnostic in a theory with broken Lorentz invariance or not. A plethora of studies have so far explored the chaotic behaviour of the holographic systems utilizing this Pole-skipping method in the black hole background \cite{Grozdanov:2017ajz, Blake:2018leo, Ahn:2019rnq, Wu:2019esr}, plasma physics \cite{Sil:2020jhr, Li:2019bgc, Amano:2022mlu}, conformal field theories \cite{Ramirez:2020qer, Haehl:2019eae, Das:2019tga}, holographic system with chiral anomaly \cite{Abbasi:2019rhy, Abbasi:2023myj}, with stringy corrections \cite{Grozdanov:2018kkt}, higher derivative corrections \cite{Natsuume:2019vcv, Wu:2019esr, Baishya:2023mgz}, brane set-up \cite{Baishya:2023ojl} etc. Moreover, this phenomenon has been extensively studied in various contexts and explored in different directions in \cite{Ceplak:2021efc, Kim:2020url, Grozdanov:2020koi, Jeong:2021zhz, Wang:2022mcq, Wang:2022xoc, Yuan:2023tft, Natsuume:2023lzy, Baishya:2023xbj}. 
\end{enumerate}We elaborate on the butterfly effect in asymptotically Lifshitz black holes by calculating butterfly velocity and the Lyapunov exponent via the above three distinct methods. Our comparative analysis delivers an exact match of the results obtained from these methods. This eventually demonstrates the equivalence between these methods for deriving quantum chaos for an asymptotically Lifshitz black hole. It is worth noting that, in \cite{Sircar:2016old}, the authors determined the scrambling time in Kruskal coordinates for the Lifshitz black hole. However, the explicit forms of the Lyapunov exponent and butterfly velocity for black holes with arbitrary anisotropy have yet to be fully explored in the context of OTOCs. Our primary objective is to calculate explicit forms of these two parameters using all the aforementioned methods. This should provide a comprehensive understanding of the chaotic properties of Lifshitz black holes. 
\par Equivalence of these methods of studying chaos was priorly found for higher-order gravity theories in \cite{Dong:2022ucb}. In this article, we focus on delivering a similar equivalence between the chosen three methods for the asymptotic Lifshitz black hole, which is a class of solutions of Einstein-Proca-type gravity theories supported by massive vector fields. It is very motivating to check whether this kind of non-relativistic theory will hold the equivalence or not.
 \par In addition, we also elucidate some of the chaotic properties, precisely, the eikonal bulk phase shift in the heavy-heavy-light-light gravitational scattering and the Lyapunov exponent, from the classical perspective. Classically, the eikonal phase is known to be related to the deflection angle of the null geodesic of a gravity background \cite{Kabat:1992tb, DAppollonio:2010krb, Bjerrum-Bohr:2014zsa}. It also finds its significance in the AdS/CFT holography, see \cite{Cornalba:2006xk, Cornalba:2006xm, Brower:2007qh} for review. In \cite{manuela}, the bulk eikonal phase shift for the heavy-heavy-light-light particle scattering in the asymptotically AdS black hole is found to be dual to the OTOCs in the Regge limit of the corresponding dual conformal field theory. Moreover, there is evidence of classical/quantum correspondence in different chaotic one-body or two-body quantum systems where the classical Lyapunov exponent can be exactly extracted from the growth rate of the OTOCs \cite{Ch_vez_Carlos_2019, PhysRevD.106.106001}. In light of such studies, we wish to understand a suitable connection between the classical bulk phase and the phase factor in the OTOCs of our chosen system. Further, we wish to propose an empirical relation between the temperature of the system and the turning point of the null geodesic for any fixed value of the anisotropy parameter $\xi$ by calculating the classical Lyapunov exponent. While deriving the eikonal phase shift, the null geodesic equation plays a dominant role in determining the impact parameter. It is defined as the ratio of conserved momenta to conserved energy. The impact parameter of the gravitational scattering crucially affects the nature of the eikonal phase. For real and imaginary eikonal phases, one must obtain elastic and completely inelastic gravitational scatterings, respectively. Here in our article, we present how the anisotropy of our chosen gravity background causes different kinds of eikonal phases that result in different types of scatterings when we gradually shift the extrema of the null geodesic from near boundary to near horizon limit.  Added to this, we verify the eikonal phase using the WKB approximation of the equation of motion of a scalar field. Specifically, we examine for a probable matching of the eikonal phase yielded from both methods. Furthermore, we wish to briefly exhibit the classical nature of the Lyapunov exponent in the anisotropic background and its consistency with the restrictions we impose on the choices of the extrema of the null geodesic to validate our approach.
 \par We organize the paper as follows. In section \ref{sec2}, we give a brief idea of the Lifshitz black hole. In section \ref{sec3}, we perform the explicit calculation of the Lyapunov exponent and butterfly velocity using all three methods as explained above. Section \ref{sec4} is devoted to the computation of the eikonal phase and Lyapunov exponent via a classical approach. We utilize the notion of geodesic stability to accomplish our purpose here. In section \ref{sec5}, we conclude with an elaborate discussion of the implications of our results for both classical and quantum chaos as well as some future scopes.
\section{Asymptotically Lifshitz black hole}\label{sec2}
In this section, we briefly revisit the black hole geometries that asymptote the non-relativistic planar Lifshitz background. Such black holes were first developed in \cite{Balasubramanian:2009rx} for $d=2,~ \xi=2$. These black holes are consistent solutions of the equations of motion of the Einstein-Proca type gravity theories with a massive vector field \cite{Taylor_2016},
\begin{equation}
\begin{gathered}
    \mathcal{S}=\frac{1}{2\kappa^2}\int d^{d+1}x\sqrt{-g}\left[\mathcal{R}-2\Lambda-\frac{1}{4}F_{\mu\nu}F^{\mu\nu}-\frac{1}{2}m^2A_{\mu}A^{\mu}\right]\,,\\
    F_{\mu\nu}=\partial_{\mu}A_{\nu}-\partial_{\nu}A_{\mu}\,.
\end{gathered}
\label{proca action}
\end{equation}Here, for any (d+1) dimensional Lifshitz black hole, the cosmological constant $\Lambda$ must be a negative quantity. Again, $F_{\mu\nu}$ is the electromagnetic field strength. The appearance of a massive vector field with mass $m$ in the above action causes the breaking of Lorentz invariance in the corresponding metric solutions. The generic form of the $(d+1)$ dimensional Lifshitz black hole solution of the above action can be written as,
\begin{equation}
\begin{gathered}
    ds^{2}=\frac{R^{2}}{z^2}\left[ -\frac{R^{2(\xi-1)}}{z^{2(\xi-1)}}f(z)dt^{2}+d\Vec{x}_{d-1}^{2}+\frac{dz^2}{f(z)}\right]\,,\\
    f(z)=1-\left(\frac{z}{z_h}\right)^{d-1+\xi}\,.
\end{gathered}
\label{Lifshitz BH 2}
\end{equation}
$R$ is the AdS radius, $z$ is the radial coordinate and $z_h$ is the black hole horizon. This geometry asymptotically reaches the planar Lifshitz spacetime
\begin{equation}
    ds^{2}=\frac{R^{2}}{z^2}\left[ -\frac{R^{2(\xi-1)}}{z^{2(\xi-1)}}dt^{2}+d\Vec{x}_{d-1}^{2}+dz^2\right]\,.
\end{equation}Akin to the planar Lifshitz spacetime, the Lifshitz black hole geometry follows anisotropic scaling along the time and space directions
\begin{equation}    t\rightarrow \Omega^\xi t,~~z\rightarrow \Omega z,~~\vec{x}\rightarrow \Omega\vec{x}\,.
\end{equation}This breaks the Lorentz invariance. Equation (\ref{Lifshitz BH 2}) describes a one-parameter family of linearly charged Lifshitz black holes that are thermodynamically stable and become extremal at the vanishing size \cite{Danielsson:2009gi}. The thermodynamics of such an asymptotically Lifshitz black hole can be reproduced from the holographic renormalization of the gravity theory represented by the action (\ref{proca action}) \cite{parkChanyong}. The legitimate dual field theory that lives on the boundary of Lifshitz black hole geometry is the finite temperature version of the non-relativistic Lifshitz field theories.  When we take 2D thermal CFT, the flat boundary that accommodates the CFT is compactified along the time direction and thus becomes a cylinder. This compactification is done by taking $t\rightarrow t+\beta$, where $\beta$ gives the circumference of the compactifying circle. Thus $\beta$ acquires the dimension of length. Now, let us consider a similar compactification in the case of the thermal LFT, where the dimension of time is [length]$^{\xi}$ due to the Lifshitz scaling. Thus, if we take $t\rightarrow t+\tilde{\beta}$ in analogy to thermal CFT for the compactification, $\tilde{\beta}$ should have the dimension of time. Therefore, on the dual bulk side, the temperature of the Lifshitz black hole becomes
\begin{equation}
    T=\frac{1}{z_h^{\xi}}\frac{d-1+\xi}{4\pi}=\frac{1}{\beta^{\xi}}\equiv\frac{1}{\tilde{\beta}}\,.
    \label{BH temperature}
\end{equation}Note that, $\tilde{\beta}$ has dimension of length. For maintaining simplicity in our further calculations, we will denote $\tilde{\beta}$ as $\beta$. 

\section{Analyses of quantum chaos} \label{sec3}
In this section, we deliver explicit calculations of the chaotic parameters using the three methods of interest. Our prime objective is to unveil similar dependencies of the chaotic parameters on the anisotropy via all these methods. 
\subsection{Entanglement wedge method}
\begin{figure*}[ht!]
\centering
\includegraphics[width=0.5\textwidth,height=4cm]{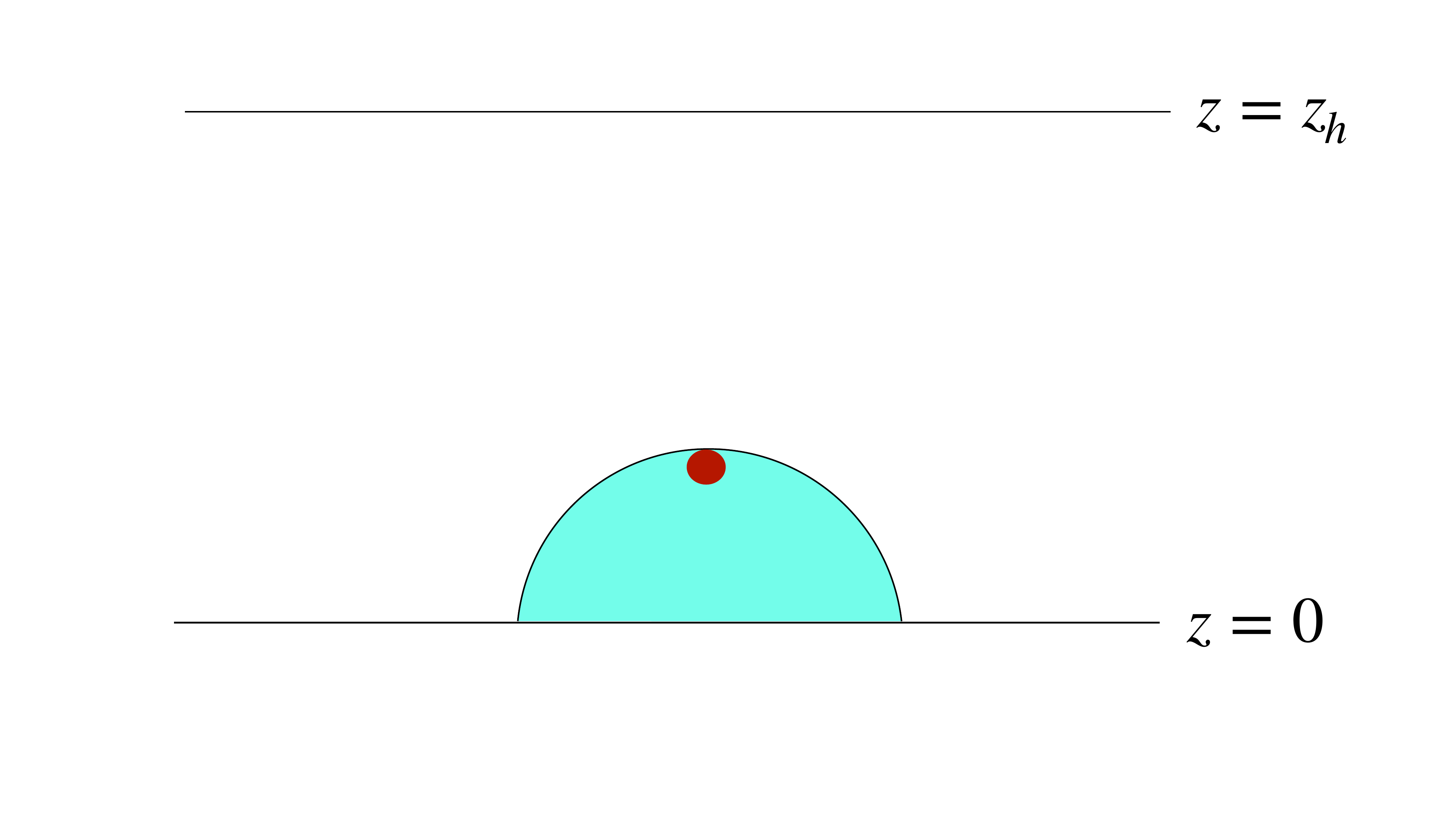}%
\includegraphics[width=0.5\textwidth,height=4 cm]{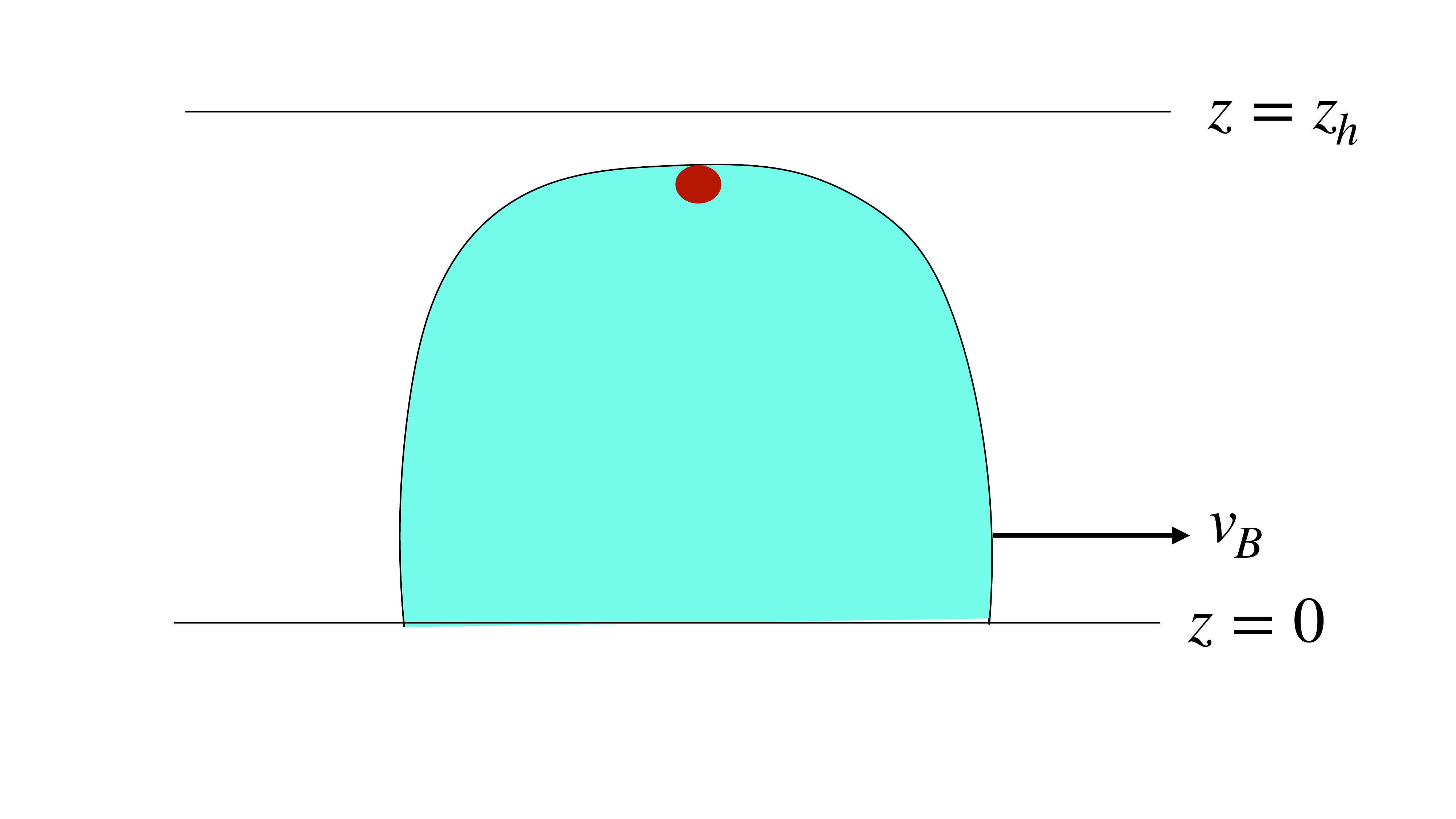}
\caption{A particle (shown by the red dot) is on the RT surface (shown by the light teal portions) bounded by the boundary region. $z=0$ is the boundary, and $z=z_h$ is set as the horizon of the black hole.}
\label{wedge}
\end{figure*}

As discussed in the introduction, in this subsection, we look into the butterfly velocity with the entanglement wedge method. For this, we need to calculate the size of the smallest boundary region whose entanglement wedge encloses the infalling particle as shown in \figurename{ \ref{wedge}}. In general, the location of the RT surface is determined by extremizing the holographic entanglement entropy functional,
\begin{equation}
    S_{\text{EE}}=2\pi\int\dd^{d-1}y\sqrt{\gamma}\,.
\end{equation}
Here, $\gamma$ is the determinant of the induced metric and $y$ includes the set of coordinates on an appropriate codimension-2 surface. 
Now, for the background \eqref{Lifshitz BH 2}, taking a constant $t$ hypersurface ($\dd t=0$) and parametrizing $z$ as $z(r)$ with $r=|x^{i}|$, we can write the induced metric as,
\begin{align}
    \gamma_{\alpha\beta}\,\dd x^{\alpha}\dd x^{\beta}=\frac{R^{2}}{z^2}\left[\left(1+\frac{z'^2}{f(z)}\right)\dd r^{2}+r^{2}\dd\Omega_{d-2}^{2}\right].
\end{align}
Note that the near-horizon and near-boundary limits for (\ref{Lifshitz BH 2}) are achieved by {$z\rightarrow z_h$} and $z\rightarrow0$ respectively. As we will work with the near-horizon region, we define the RT surface in the {$z=z_h$} by,

\begin{equation}\label{rtsurface}
    z(r)=z_h-\epsilon\,y(r)^{2},
\end{equation} where we parametrize the near-horizon limit({$z=z_h$}) with a new function $y(r)$. The coefficient $\epsilon\,(>0)$ is a very small number, and the function $y(r)$ is the RT profile. We deduce the RT profile from the RT equation by performing a Taylor series expansion around $\epsilon=0$. Now, expanding the induced metric up to order $\epsilon$, we get,
\begin{widetext}
    \begin{equation}
     \gamma_{\alpha\beta}\,\dd x^{\alpha}\dd x^{\beta}=R^{2}\left[\left(\frac{4\,\epsilon^{2}y^{2}y'^2}{(z_h-\epsilon\,y^2)^{2}f_{1}(z_h-z)}+\frac{1}{(z_h-\epsilon\,y^2)^2}\right)\dd r^{2}\right. \\ \left.+\frac{r^2}{(z_h-\epsilon\,y^2)^2}\dd\Omega_{d-2}^2\right]\,,\nn\\
     \end{equation}
     \begin{equation}
     =\frac{R^{2}}{z_h^2}\left[\left(1+\frac{2\epsilon}{z_h} (y^{2}+\frac{2}{f_1}y'^2)\right)\dd r^2+r^{2}(1+\frac{2\epsilon}{z_h}y^2)\dd\Omega_{d-2}^2+\mathcal{O}(\epsilon^2)\right],
\end{equation}
\end{widetext}

 where we Taylor expand the factor $f(z)$ near the horizon up to the first order as $f(z)\approx f_{1}(z_h-z)$. The determinant of the induced metric is given by,
 \begin{align}
     \sqrt{\gamma}=\frac{R^{2}r^{d-2}}{{z_h^{d-1}}}\left[1+\frac{\epsilon}{{z_h}}\left((d-1)y(r)^{2}+\frac{2}{f_1}{z_h}y'(r)^{2}\right)\right]\nonumber\\+\mathcal{O}(\epsilon^2)\,.
          \label{gamma}
 \end{align}
Now, we write the RT equation by varying (\ref{gamma}) with respect to $y(r)$ as
\begin{equation}
    (d-1)y(r)-\frac{2{z_h}}{f_1}\left(y''(r)+(d-2)\frac{y'(r)}{r}\right)=0\,.
\end{equation}where we consider the terms only up to order $\epsilon$.
 Solving this second-order differential equation, we get,
 \begin{equation}
y(r)=r^{-n}\left[\text{J}_{n}(\mu r)+\text{Y}_{n}(\mu r)\right]\,,
\label{y equation}
 \end{equation}
 where $\text{J}_{n}$ and $\text{Y}_{n}$ are the Bessel functions of first kind and second kind, respectively. In the equation (\ref{y equation}), 
 \begin{equation}
     n=\frac{1}{2}(d-3),\hspace{1cm}|\mu|=\sqrt{\frac{(d-1)(d-1+\xi)}{2{z_h^2}}}\,.
 \end{equation}
Here, the scaling parameter $\mu$ behaves as the momentum. From the eq. \eqref{rtsurface}, we  see that the function $y(r)$ takes the form of $0$  near the horizon ${z_h}$. In this vicinity, both Bessel functions behave as $\sim\mu^n$. In \cite{Dong:2022ucb}, the authors have calculated the butterfly velocity with an exponential ansatz. In our work, we prescribe a more general result for calculating butterfly velocity without taking any specific ansatz.
\par The surface enclosing the particle approaches the horizon at a constant speed, termed as butterfly velocity ($v_B$). We assume that, at each point in time, the tip of the RT surface touches the particle that is visualised by demanding $y(r=0,t)\sim e^{-\frac{2\pi}{\beta}t}$. 
Therefore, with the value of $\mu$, we can calculate the butterfly velocity as,
\begin{align}
    v_{B}&=\frac{2\pi}{\beta|\mu|}=\frac{2\sqrt{2}\pi {z_h}}{\beta\sqrt{(d-1)(d-1+\xi)}}\,.
\end{align}Eventually, at the near-horizon limit, we find,
\begin{equation}\label{butter_entan}
     v_{B}={z_h^{1-\xi}}\sqrt{\frac{d-1+\xi}{2(d-1)}}\,.
\end{equation}
In the $\xi\rightarrow 1 $ limit, we get the planar black hole result $v_{B}=\sqrt{\frac{d}{2(d-1)}}$. So, the anisotropy affects the butterfly velocity nontrivially. Butterfly velocity characterizes the propagation of chaos in a local system. In a series of preceding literature \cite{Giataganas:2014hma, Bai:2014wpa, Giataganas:2017koz}, it is robustly established that the chaotic features of the Lifshitz invariant systems increase for large $\xi$ values, i.e., for large anisotropy. The dynamical critical exponent $\xi$ in the Lifshitz class of gravity backgrounds introduces anisotropic scaling along the temporal direction and hence breaks Lorentz symmetry. Therefore, depending on the values of $\xi$, there appears certain deformation of the spacetime geometry from the usual conformal AdS.  This causes change in the shape of the geodesic of an infalling particle inside the bulk. For Lifshitz black hole, the position of the horizon becomes more prominent with the increase in $\xi$ and the geodesics of infalling particles are deviated further away from each other. Thus, with increasing anisotropy, the anisotropic deformation in the Lifshitz geometry causes chaotic growth in the associated gravity theory as well as its holographic dual field theory. It has been found that the chaotic evolution happens monotonically with larger values of $\xi$. We get similar scenario of monotonically increasing $v_B$  with the increase in the anisotropy parameter in all of our chosen methods.
From the expression of $v_B$ in (\ref{butter_entan}), it is obvious that $v_B$ increases monotonically with $\xi$ for any fixed dimension of the bulk Lifshitz black hole. Furthermore, we can substitute for $z_h$ in terms of $T$ in the equation (\ref{butter_entan}) by using (\ref{BH temperature}) so that
\begin{equation}\label{butterfly_temp}
    v_B=(4\pi T)^{1-\frac{1}{\xi}}\frac{1}{\sqrt{2(d-1)}}\left(d-1+\xi\right)^{\frac{2-\xi}{2\xi}}
\end{equation}Therefore, for any fixed $\xi>1$, the butterfly velocity is found to be a monotonically increasing function of the Lifshitz black hole temperature $T$. However, in the absence of anisotropy, i.e., for $\xi=1$, $v_B$ becomes independent of the temperature, which is consistent with the case for the usual asymptotically AdS black hole.

\subsection{Out-of-time-ordered correlators}\label{otoc_subsec}
In this subsection, we will study the out-of-time-ordered correlator (OTOC) for thermal Lifshitz field theory dual to a linearly charged asymptotically Lifshitz black hole with arbitrary anisotropy. It is convenient to study the OTOC in the Kruskal-Szekeres form of the bulk metric, which smoothly covers the globally extended spacetime. In Kruskal geometry, the OTOC can be interpreted as the amplitude of the two-particle gravitational scattering where the particles move along the two horizons \cite{Shenker:2014cwa, Maldacena:2015waa, Shenker:2013yza, Roberts:2014isa}. We will use the generic expression of the OTOC
\begin{equation}
\langle\hat{W}(t_2,x_2)\hat{V}(t_1,x_1)\hat{W}(t_2,x_2)\hat{V}(t_1,x_1)\rangle_{\beta}\,,
\label{otoc1}
\end{equation}compute the OTOC in the specific dual-field theory. Here we consider $t_2-t_1>>\beta$. The particles that participate in the gravitational scattering process are the operators $\hat{W}$ and $\hat{V}$, which are inserted in the dual thermofield double (TFD) states. Though the quantity \eqref{otoc1} is one-sided, our whole computation of the OTOC has been done in a two-sided geometry. So far, the OTOC via the gravitational scattering process has been explored with various black hole backgrounds \cite{Shenker:2014cwa, Maldacena:2015waa, Shenker:2013yza, Roberts:2014isa, Ahn:2019rnq, Blake:2021hjj, Fischler:2018kwt, Jahnke:2019gxr, Mezei:2019dfv, Reynolds:2016pmi, Poojary:2018esz, Banerjee:2019vff, Saha:2024bpt}. In the Lifshitz background, the authors of \cite{Sircar:2016old} calculated the scrambling time, though the background they worked with is slightly different from ours. Moreover, OTOCs for a generic holographic bulk including specific classes of anisotropic black holes are also been derived analytically in \cite{Giataganas:2017koz, Jahnke:2017iwi}. Here, we provide an explicit derivation of the double-sided Kruskal-Szekeres version of the Lifshitz black hole and analytically develop the OTOCs. We will show that our constructed OTOCs are consistent with those found for the generic holographic dual gravity theories. We use these OTOCs to subsequently calculate the Lyapunov exponent and butterfly velocity in the two-sided geometry, which yield interesting results.
\subsubsection{Kruskal extension}
 To study the out-of-time-ordered correlators, let us first construct the Kruskal extension of the (2+1) dimensional Lifshitz black hole. We start with the metric \eqref{Lifshitz BH 2} while replacing $z=1/r$ and considering the AdS radius to be 1 for simplicity,
  \begin{equation}
    \dd s^2=-r^{2\xi}f(r)\dd t^2+\frac{1}{r^2}\frac{\dd r^2}{f(r)}+r^2\dd{x}^2\,,
    \label{Lifshitz BH}
\end{equation} The blackening factor $f(r)$ is then given by, 
\begin{equation}
    f(r)=1-\left(\frac{r_h}{r}\right)^{1+\xi}\,.
\end{equation}
$r$ is the radial coordinate and $r_h$ is the black hole horizon in the emblackening factor $f(r)$. In \cite{Alvarez:2014pra, Zhu_2020}, geometries of nonlinearly charged Lifshitz black holes with one and two horizons are obtained with the Eddington-Finkelstein coordinates. In this case, nonlinearity appears in the blackening factors and is characterized by some extra parameters along with the dynamical exponent $\xi$. Single horizons and double horizons for such black holes are obtained by using specific parametric conditions. However, for the metric in (\ref{Lifshitz BH}), we will construct an explicit form of the linearly charged thermal Lifshitz black hole with double horizons by using the Kruskal-Szekeres coordinates. To work in these coordinates, we transform,
\begin{equation}
    u=e^{\frac{\alpha}{2}(r_*-t)},~~v=e^{\frac{\alpha}{2}(r_*+t)}\,,
    \label{Kruskal}
\end{equation}
where $\alpha$ is a function of $r_h$ that relates with the finite temperature $T$ of the black hole as $\alpha\equiv 2\pi T$. The tortoise coordinates $r_*$ play a crucial role in defining the smooth horizon in the ($u,v$) coordinate. We define, 
\begin{equation}
    r_*(r)=\int\frac{\dd r}{{r}^{1+\xi}f(r)}=\int\frac{\dd r}{{r}^{1+\xi}\left[1-\left(\frac{r_h}{r}\right)^{1+\xi}\right]}\,.
    \label{r*}
\end{equation}  
Integrating equation (\ref{r*}), we get, 
\begin{equation}
    r_{*}(r)\approx\frac{r^{-\xi}}{\xi(1+\xi)}\,{_{2}}F_{1}\left(1,-\xi,1-\xi,\frac{r}{r_h}\right)\,.
    \label{r1*}
\end{equation}
We consider $f(r)=(r-r_h)f'(r_h)$ at the near-horizon limit during the integration. The limit $r\rightarrow r_h$ further reduces (\ref{r1*}) to 
\begin{equation}
    r_{*}\approx \frac{1}{r_{h}^{\xi}(1+\xi)}\log\left(\frac{r-r_h}{r_h}\right)\,.
\end{equation}
As there is a relation between ($u,v$) coordinate and $r_*$ (from \eqref{Kruskal}), we can easily write,
\begin{equation}\label{uvrel}
    uv=e^{\alpha r_*}\approx\left(\frac{r-r_h}{r_h}\right)\,.
\end{equation}
This equation allows us to claim that, in the Kruskal geometry, $uv=0$ is defined as the horizon while singularity is at $uv=1$ and both left/right boundaries are located at $uv=-1$. Writing the metric \eqref{Lifshitz BH} in terms of the null coordinates $u$ and $v$, we achieve
\begin{equation}
\begin{split}
    \dd s^2=&\frac{4r(u,v)^{2\xi}}{\alpha^2 uv}\,f(r(u,v))\dd u\,\dd v+r(u,v)^2 \,\dd x^2\\&=2A(u,v)\dd u\,\dd v+B(u,v)\dd x^2\,.
    \end{split}
\end{equation}This unperturbed background obeys Einstein's equation, 
\begin{equation}\label{einstein}
    E_{\mu\nu}=\kappa T_{\mu\nu}\,.
\end{equation}
Here, $E_{\mu\nu}$ is the Einstein tensor, $\kappa=8\pi G_{N}$ is a constant related to the Newton's constant $G_N$ and $T_{\mu\nu}$ is the stress-energy tensor
\begin{equation}
    T_{0}^{\text{matter}}=2 T_{uv}\dd u\dd v+T_{uu}\dd u^{2}+T_{vv}\dd v^{2}+T_{xx}\dd x^{2}\,.
\end{equation}
It is worth mentioning that this stress tensor accounts for the cosmological constant and is consistent with the Ricci tensor of the unperturbed background. For the unperturbed Einstein equation, solving $T_{uu},T_{vv}$ and $T_{uv}$, we get,
\begin{widetext}
\begin{subequations}
\begin{align}
    &T_{uu}=\frac{1}{4A(u,v)B(u,v)^2}\left[A(u,v)\partial_{u}B(u,v)^{2}\right.  \left.+2B(u,v)\left(\partial_{u}A(u,v)\partial_{u}B(u,v)-A(u,v)\partial_{u}^{2}B(u,v)\right)\right]\,,\\&
    T_{uv}=\frac{1}{4B(u,v)^2}\left[2B(u,v)\partial_{u}\partial_{v}B(u,v)-\partial_{v}B(u,v)\partial_{u}B(u,v)\right]\,,\\&
    T_{vv}=\frac{1}{4A(u,v)B(u,v)^2}\left[A(u,v)\partial_{v}B(u,v)^{2}+2B(u,v)\left(\partial_{v}A(u,v)\partial_{v}B(u,v)\right.\right. \left.\left.-A(u,v)\partial_{v}^{2}B(u,v)\right)\right]\,.
\end{align}
\end{subequations}
\end{widetext}

\subsubsection{Gravitational backreaction due to shock wave}
\label{subsubsection }
Followed by the construction of a two-sided 2+1D Lifshitz black hole geometry, we now introduce a null pulse of energy localized along the $v=0$ horizon and study the resulting back-reacted metric. In the bulk-boundary picture, this means we insert an operator in the boundary thermal state at some past time $t$ and let it evolve. Inserting an operator in the boundary leads to an infalling perturbation close to the boundary. The energy of the perturbation on a constant time slice exponentially increases with time $t$ \cite{Shenker:2013pqa}. Thus, we may approximate this perturbation as a null pulse of energy $E$ localized at the $v=0$ horizon along the $u$-direction. Let us consider the back-reaction of this null pulse in the two-sided geometry.
\begin{widetext}
\usetikzlibrary { decorations.pathmorphing, decorations.pathreplacing, decorations.shapes}
\vspace{0.5cm}
\begin{figure}[ht]
    \centering
    \begin{tikzpicture}[thick]\hspace{0.5cm}
  \draw[decorate,decoration={coil,aspect=0, segment length=4pt, amplitude=2pt}]  (3,1) to [out=10,in=170] (8,1);
  \draw[decorate,decoration={coil,aspect=0,  segment length=4pt, amplitude=2pt}]           (3,5)  to[out=-10,in=190] (8,5);
  \draw [dashed]   (3,1)  -- (8,5) node[pos=0.25, sloped, below, rotate=2] {$u=0$};
   \draw [dashed]   (3,5)  -- (8,1)node[pos=0.75, sloped,above, rotate=2] {$v=0$};
    \draw   (3,1)  -- (3,5);
     \draw   (8,1)  -- (8,5);
     \draw[->] (5.4,3.3) -- (4.4,4.1) node[above] {$u$};
      \draw[->] (5.55,3.3) -- (6.6,4.1)node[above] {$v$};
      \draw[decorate,decoration={coil,aspect=0, segment length=4pt, amplitude=2pt}]  (11,1) to [out=10,in=170] (16,1);
  \draw[decorate,decoration={coil,aspect=0,  segment length=4pt, amplitude=2pt}]           (11,5)  to[out=-10,in=190] (16,5);
   \draw [dashed]   (11,5)  -- (16,1);
  \draw[red,decorate,decoration={zigzag,aspect=0,  segment length=4pt, amplitude=2pt}]           (16,1.1) -- (11,5);
    \draw   (11,1)  -- (11,5);
     \draw   (16,1)  -- (16,5);
     \draw[->,blue] (11,2) -- (12.9,3.5);
      \draw[->,blue] (13.8,2.9) -- (16,4.9);
      \begin{scope}
    \clip (11,5) to[out=-10,in=190] (16,5) -- (16,5) -- (16,1) -- (11,1) -- (11,5);
    \fill[green!25,nearly transparent] (11,5) -- (16,1) -- (16,5) -- cycle;
  \end{scope}
\end{tikzpicture}
    \caption{\textit{Left:} Penrose diagram of two-sided geometry. \textit{Right:} A shock wave (\textcolor{red}{red zigzag line}) localised along $v=0$ causes a discontinuity of the $u=0$ horizon (the blue arrow)}
    \label{fig:enter-label}
\end{figure}
\end{widetext}
We consider the following form of the stress-tensor localised at $v=0$,
\begin{equation}\label{shockeq}
    T_{vv}^{\text{shock}}=E\exp(\frac{2\pi t}{\beta})\delta(v)\delta(x)\,.
\end{equation}
The shock wave localised at $v=0$ horizon indeed splits the causal past (white) and future (green) of $v$. The metric in the green region changes while in the white region, it remains the same. In the Penrose diagram, one can clearly observe the shift along the $v$-direction. We can write the form of this shift as
\begin{align}
    &\bar{u}=u+\Theta(v)\eta(u,v,x)\rightarrow \dd{u}=\dd\bar{u}-\eta(u,v,x)\delta(v)\dd v\,,\\
    &\bar{v}=v\,,\hspace{2cm}\bar{x}=x\,.
\end{align}
Our purpose is to acquire the form of the function $\eta(u,v,x)$ -- the shockwave profile. The Heaviside step function $\Theta(v)$ ensures that only the causal future of the pulse is affected by its presence. In terms of the new coordinates ($\bar{u},\bar{v},\bar{x}$), we can write the metric as,
\begin{equation}
    \dd s^2=2 A(\bar{u},\bar{v})\dd\bar{u}\dd\bar{v}+B(\bar{u},\bar{v})\dd\bar{x}^{2}-2A(\bar{u},\bar{v})\eta(u,v,x)\delta(\bar{v})\dd\bar{v}^2\,.
\end{equation}
With such a back-reacted form of the metric, we need to reach a form of $\eta(u,v,x)$ which will obey eq.\eqref{einstein}. In the above metric,
\begin{align}
    A(\bar{u},\bar{v})=\frac{2r(\bar{u},\bar{v})^{2\xi}}{\alpha^{2}\bar{u}\bar{v}
    }f(r(\bar{u},\bar{v})),~~B(\bar{u},\bar{v})=r(\bar{u},\bar{v})^2\,,
\end{align}
where we express the radial coordinate $r$ in terms of ($u,v$). Now, for the perturbed background, we wish to understand the stress tensor. The stress tensor for the matter part reads as
\begin{equation}\label{matter}
\begin{split}
&T_{\mu\nu}^{\text{matter}}=2\left[T_{\bar{u}\bar{v}}-2T_{\bar{u}\bar{u}}\eta(\bar{u},\bar{v},x)\delta(\bar{v}\right]\dd\bar{u}\dd\bar{v}\\&+\left[T_{\bar{v}\bar{v}}+T_{\bar{u}\bar{u}}\eta(\bar{u},\bar{v},x)^{2}\delta(\bar{v})^{2} -2T_{\bar{u}\bar{v}}\eta(\bar{u},\bar{v},x)\delta(\bar{v})\right]\dd\bar{v}^2\\&+T_{\bar{u}\bar{u}}\dd \bar{u}^2+T_{\bar{x}\bar{x}}\dd\bar{x}^2\,.
\end{split}
\end{equation}
We will drop the bar convention for further analyses so that the result looks simpler. Now, let us solve the back-reacted Einstein's equation, which assumes the form
\begin{equation}
    E_{\mu\nu}=\kappa\left(T_{\mu\nu}^{\text{matter}}+T_{\mu\nu}^{\text{shock}}\right)\,.
\end{equation}
The expressions of $T_{\mu\nu}^{\text{matter}}$ and $T_{\mu\nu}^{\text{shock}}$ are given by \eqref{matter} and \eqref{shockeq} respectively. Solving the $vv$ component of the perturbed Einstein equation yields
\begin{equation}
\begin{split}
      \left[\partial_{x}^{2}-\frac{1}{A(u,v)}\partial_{u}\partial_{v}B(u,v)\right]\eta(u,v,x)-\\ \frac{8\pi G_{N} E\,B(u,v)}{A(u,v)} e^{2\pi t/\beta}\delta(v)\delta(x)=0\,,
       \end{split}
      \end{equation}
      \begin{equation}
    \Rightarrow\left[\partial_{x}^{2}-\mathcal{M}(u,v)^2\right]\eta(u,v,x)=e^{\frac{2\pi}{\beta}(t-t_*)}\delta(v)\delta(x)\label{shock1}\,.
\end{equation}
Where,\begin{equation}\label{scram}
    \begin{split}
    \mathcal{M}(u,v)=\sqrt{\frac{\partial_{u}\partial_{v}B(u,v)}{A(u,v)}}\, ,\\\hspace{.5cm}t_{*}(u,v)=\frac{\beta}{2\pi}\log\left(\frac{A(u,v)}{8\pi G_{N} E\,B(u,v)}\right)\,.
     \end{split}
\end{equation}
We will discuss the physical meaning of the function $t_{*}(u,v)$ in what follows. After implementing the values of $A(u,v)$ and $B(u,v)$ to $t_*(u,v)$, it turns out to be a constant. It is evident that the shockwave profile $\eta(u,v,x)$ obeys an ODE which can be solved easily. While writing the above equation, we assume $T_{uu}=0$ near the $v=0$ horizon \cite{Fischler:2018kwt}. 
Near the horizon ($v=0$ or $r=r_h$),
 \begin{equation}
     A(u,v)|_{v=0}=\frac{8}{1+\xi},\hspace{0.5cm}\partial_{u}\partial_{v}B(u,v)|_{v=0}=4r_{h}^2\,.
 \end{equation}During the differentiation of $B(u,v)$ with respect to $u$ and $v$, we use the relation \eqref{uvrel} and performe a chain rule differentiation with $r_*$. Now, implementing the above two expressions into \eqref{scram}, we get,
\begin{equation}
\begin{split}
    &\mathcal{M}(u,v)|_{v=0}=\sqrt{\frac{{r_h^2}(1+\xi)}{2}}\,,\\&t_{*}=\frac{\beta}{2\pi}\log\left(\frac{1}{\pi G_{N} E(1+\xi)r_h^2}\right)\,,
    \label{M and t}
\end{split}
\end{equation}  
where we consider $r_h=1$. The constant $t_*$ is called the scrambling time, which depends on the anisotropy index $\xi$. Near the horizon, the equation \eqref{shock1} eventually becomes,
\begin{equation}
     \left[\partial_{x}^{2}-\frac{{r_h^2}(1+\xi)}{2}\right]\eta(x)=e^{\frac{2\pi}{\beta}(t-t_*)}\delta(x)\,.
\end{equation}
We first solve the homogeneous equation and achieve,
\begin{equation}
    \eta(x)=\begin{cases}
        c_{1}e^{\mathcal{M}x}+c_{2}e^{-\mathcal{M}x}\quad \text{for}~~x>0\,,\\
         c_{3}e^{\mathcal{M}x}+c_{4}e^{-\mathcal{M}x}\quad \text{for}~~x<0\,.
    \end{cases}
\end{equation}
Now, to solve the non-homogeneous equation, we check the discontinuity of the first derivative of the solution and get,
\begin{equation}\label{discon}
    \eta'(\epsilon)-\eta'(-\epsilon)=e^{\frac{2\pi}{\beta}(t-t_*)}\,,
\end{equation}
where, $\epsilon$ is a small parameter $\epsilon\rightarrow 0$. Imposing the continuity of the solution at $x=0$ along with the discontinuity equation \eqref{discon}, we get a relation between the coefficients as,
\begin{equation}
    c_{1}-c_{3}=c_{4}-c_{2}=\frac{4\sqrt{2}e^{\frac{2\pi}{\beta}(t-t_*)}}{r_h(1+\xi)^{3/2}}\,.
\end{equation}
Let us choose $c_2=c_3=0$. With this, the final solution for the shock is,
\begin{equation}
    \eta(t,x)=\frac{4\sqrt{2}}{r_h((1+\xi)^{3/2}}e^{\frac{2\pi}{\beta}(t-t_*-\frac{\beta\mathcal{M}}{2\pi}x)}\,.
    \label{otoc}
\end{equation}
This shockwave profile completely determines the four-point out-of-time ordered function. We can identify the commutator $C(t,x)$ in equation (\ref{commutator}) with the shockwave profile $\eta(t,x)$ and write,
\begin{equation}
    \lambda_{L}=2\pi T\,,\hspace{2cm}v_{B}={r_{h}^{\xi-1}}\sqrt{\frac{1+\xi}{2}}\,.
\end{equation} This Lyapunov exponent agrees with our results of the entanglement wedge method. For the butterfly velocity, this result is found to be the same as obtained from the entanglement wedge result for $d=2$ in \eqref{butter_entan}. Thus, we can claim that these two methods are congruent for a reliable study of chaos in the Lifshitz background. 

\subsection{Pole-skipping}\label{pole_skipping}
In this subsection, we wish to study the retarded energy density correlation function $G^{\text{R}}_{T_{00}T_{00}} (\omega, k)$ near the pole-skipping points \eqref{poleskip}. The correlation function can be understood by perturbing the gravitational background \eqref{Lifshitz BH 2} and imposing ingoing boundary conditions at the black hole horizon. To perform the pole-skipping analysis, it is customary to work with Eddington-Finkelstein coordinates, defined by
\begin{equation}
    v=t+r_{*},\hspace{2cm}\frac{\dd r_*}{\dd r}=\frac{1}{r^{1+\xi}f(r)}\,,
\end{equation}
where $v$ is the null coordinate and $r_*$ is the tortoise coordinate. With the above coordinate system, the background metric \eqref{Lifshitz BH 2} is transformed into
\begin{equation}
    \dd s^{2}=-r^{2\xi}f(r)\dd v^{2}+2r^{\xi-1}\dd v\dd r+r^{2}\dd x^2\,.
\end{equation}
For our current analysis, we choose the gravity background as the (2+1)-D Lifshitz black hole.  Now, to calculate the energy density correlation function, we must perturb the background metric, specifically the sound modes (in other words, the longitudinal modes). We perturb the background metric as,
\begin{equation}
    g_{\mu\nu}\rightarrow g_{\mu\nu}+\delta g_{\mu\nu}(r)e^{-i\omega v+ikx}\,.
\end{equation}Here we perform a Fourier transformation to the perturbed mode and choose the wavenumber $k$ along the $x$-direction. The sound modes thus become \begin {equation}
    \delta g_{vv},\delta g_{vx},\delta g_{vr},\delta g_{rr},\delta g_{rx},\delta g_{xx}\,.
\end{equation} 
As the wave moves along the $x$ - direction, the sound modes are designated by the modes only along the $(v,r, x)$ directions. At this stage, we get some redundant modes imposing radial gauge condition $\delta g_{r\mu}=0$ and tracelessness condition $g^{\mu\nu}\delta g_{\mu\nu}=0$. Hence, we are left with
\begin{equation}
    \delta g_{vv},\delta g_{vx}\,.
\end{equation}
The retarded energy density Green's function is governed by the perturbed equations which are regular at the horizon in the ingoing EF coordinates. So, we Taylor expand the modes near the horizon as,
\begin{equation}\label{regu}
    \delta g_{\mu\nu}(r)= \delta g_{\mu\nu}^{(0)}+(r-r_h) \delta g_{\mu\nu}^{(1)}+...\,,
\end{equation}
Perturbing the sound modes of the background and imposing the regularity condition \eqref{regu} near the black hole horizon, we can achieve the expanded form of Einstein's equations near the horizon. We wish to verify whether we get exactly similar results of chaos as obtained in the previous subsections. For the field $\delta g_{vv}$, the associated equation of motion is $\delta E_{vv}=\delta T_{vv}$. However, the perturbation to stress tensor $\delta T_{vv}$ does not vanish necessarily, while $\delta T^{r}_{v}$ vanishes at the horizon \cite{Blake:2018leo, Wang:2022mcq}. Keeping this in mind, we set $\delta T^{r}_{v}=0$ to calculate the pole-skipping point connected to chaos. Expanding this equation near the black hole horizon, we get,

{\begin{equation}
\begin{split}
   \left[k^2-i \omega r_h^{2-\xi }\right]\delta g_{vv}^{(0)}+k\left[-ir_{h}^{\xi}(1+\xi)+2\omega\right]\delta g_{vx}^{(0)} =0\,.
   \end{split}
\end{equation}}
This equation represents a constraint relation between the near-horizon coefficients. At some specific value $\omega=\omega_{*}=\frac{i}{2}r_{h}^{\xi}(1+\xi)$, coefficient of $\delta g_{vx}^{(0)}$ is 0. So, at this specific value of $\omega$, we get,

\begin{equation}
    \left[k^{2}+\frac{1}{2}r_{h}^{2}\left(1+\xi\right)\right]\delta g_{vv}^{(0)}=0\,.
\end{equation}

For general $k$, this equation gives rise to $\delta g_{vv}^{(0)}=0$. However, at $k=k_{*}$, this equation is automatically satisfied. That means, at these specific values of $\omega$ and $k$, $\delta T^{r}_{v}$ is automatically zero. Consequently, they will not impose any constraint on the near-horizon coefficients $\delta g_{vv}^{(0)}\,$ and $\delta g_{vx}^{(0)}$. The frequency ($\omega_*$) and momentum ($k_*$) define the pole-skipping point. At the pole-skipping point, we thus get, 

\begin{equation}
      \omega_{*}=\frac{i}{2}r_{h}^{\xi}(1+\xi)\,,\hspace{1cm} k_{*}^2=-\frac{1}{2}r_{h}^{2}\left(1+\xi\right)\,.
\end{equation}
Writing $r_h$ in terms of temperature as $r_{h}=\left(\frac{4\pi T}{1+\xi}\right)^{1/\xi}$, we can express $\omega$ and $k$ in terms of the temperature of the black hole. One can therefore calculate the Lyapunov exponent and butterfly velocity from \eqref{poleskip} as,

  \begin{equation}\label{chaos_PS}
      \lambda_{L}=2\pi T\,,\hspace{2cm}v_{B}={r_{h}^{\xi-1}}\sqrt{\frac{1+\xi}{2}} \,.
  \end{equation} respectively.
Note that the butterfly velocity ($v_B$) is a positive real quantity. So, we will take the absolute value of momentum while calculating $v_B$, i.e. $v_{B}=\frac{|\omega_*|}{|k_*|}$. From the above expression, it is obvious that both the Lyapunov exponent and butterfly velocity assume the same forms as those achieved in the previous subsections. Henceforth, we claim that all three methods to compute the chaotic features are equivalently reliable. These methods also give consistent results in higher orders. One can check it explicitly. 
\par One can note that, for the class of hyperscale violating Lifshitz backgrounds, the metric components in the near-horizon limit involve a hyperscale violating parameter $\theta$ on top of Lifshitz scaling. This causes a nontrivial appearance of this parameter in different chaotic measures. Thus, using the above three computational methods, one can expect equivalent $\theta$-dependent expressions of $v_B$ and $\lambda_L$ for the hyperscale violating class of Lifshitz gravity background. Moreover, it has been found in \cite{Giataganas:2017koz} that the Lifshitz geometry can be obtained in the IR zone when the AdS background in UV undergoes a holographic renormalisation group (RG) flow. Thus, the conformality of the UV theory is broken in the IR due to the appearance of $\xi\neq 1$. Subsequently, a distinct confinement/deconfinement like critical phase transition scenario occurs when $\xi$ attains the values such that $\xi\neq 1$. Therefore, the anisotropy parameter $\xi$ is somewhat suggestive of the critical phase transition temperature during the RG flow. As in our case, $v_B$ and $\lambda_L$ derived from each of the aforementioned methods are nontrivial functions of $\xi$, and this anisotropic index is related to the black hole temperature $T$ via \eqref{BH temperature}. Added to this, it has been observed for each of the above studies that $v_B$ and $\lambda_L$ pick the expressions of those for AdS black hole when $\xi=1$. This makes us claim that a critical phase transition under the holographic RG flow can be traced out from the chaotic features analysed by using all these equivalent frameworks. 
\section{Analysis of classical chaos}\label{sec4}
Here we perform a parallel survey of some of the chaos properties, to be precise, eikonal phase shift and Lyapunov exponent, at the classical level. 
 We will show that the bulk phase shift may be written in terms of the series expanded exponential growth rate of the quantum OTOCs up to its leading order. Similarly to the conformal case \cite{Kim_2021}, 
the bulk phase shift happens to be a hypergeometric function. Nevertheless, it nontrivially depends on the anisotropy parameter $\xi$ for the asymptotically Lifshitz black hole. Also, we will compute the Lyapunov exponent at the classical level and check any possible relevance of the same with the $\lambda_L$ obtained by using the three different methods for quantum chaos. 
 \subsection{Null geodesic}
\label{null}
The anatomy of the null geodesics for our specified background and the corresponding turning points adequately serve the purpose of calculating both the eikonal bulk phase shift as well as the Lyapunov exponent in the classical sense. In our chosen gravity background, the presence of an arbitrary anisotropy parameter $\xi$ leads to an arbitrary number of positive real roots for the null geodesics, depending on the value of $\xi$. Moreover, all of those roots contain a nontrivial dependency on $z_h$ and $\gamma$. As a prerequisite for the classical analysis, here we present a quantitative evaluation of the null geodesics and their turning points in our framework. The bulk phase shift of a particle described by a plane wave can be written as \cite{manuela, Parnachev:2020zbr} 
\begin{equation}
    \delta\equiv -p.(\Delta x)\,,
\end{equation}where, $p$ denotes the components of the conserved momentum vector and $\Delta x$ represents the deflections in the corresponding bulk directions. For our 2+1D metric (\ref{Lifshitz BH 2}), the conserved energy and momentum are given by
\begin{equation}
    p_t=\frac{1}{z^{2\xi}}f(z)\frac{\partial t}{\partial s}\,,~~p_x=\frac{1}{z^2}\frac{\partial x}{\partial s}\,,
    \label{eom}
\end{equation} $s$ being the intrinsic affine parameter. These conserved quantities correspond to the killing vectors $\partial_t$ and $\partial_x$ of our chosen metric during a heavy-heavy-light-light 2-particle scattering. It is convenient to consider one of the particles to be highly boosted, i.e., it consists of large conserved momentum. We introduce a parameter $\gamma=\frac{p_x}{p_t}$ which serves as the impact parameter of the corresponding scattering process. The null geodesic equation is given by
\begin{equation}
    g_{\mu\nu}\dot{x}^{\mu}\dot{x}^{\nu}=0\,,
\end{equation} which yields
\begin{equation}
    \dot{z}^2=(p_t)^2z^{2\xi+2}-(p_x)^2 z^4f(z)\,.
    \label{geodesic}
\end{equation}At the turning point $z=z_0$, $z$ attains extreme value for which we can set $\dot{z}|_{z=z_0}=0$. Hence, the values of turning points $z_0$ should satisfy the relation
\begin{equation}
    z_0^{2\xi-2}+\frac{\gamma^2}{z_h^{\xi+1}}z_0^{\xi+1}-\gamma^2=0\,.
    \label{geodesic 2}
\end{equation}The equation (\ref{geodesic 2}) is a polynomial of either integer order or fractional order, depending on the values of $\xi$. Therefore, the valid roots of (\ref{geodesic 2}) for our approach should be described by a nontrivial function of $z_h$ and $\gamma$. This function must attain positive real values that lie within the singularity and the boundary of the bulk geometry. 
At this stage, some crucial remarks on the turning point follow immediately:
\begin{itemize}
    \item Let us take the assumption that the energy $p_t$ is much smaller than the momentum $p_x$, i.e., $\gamma\gg 1$. With this assumption, taking $z_0=1$ equation \eqref{geodesic 2} gives, 
    \begin{equation}
        \gamma=\sqrt{\frac{z_h^{\xi+1}}{z_h^{\xi+1}-1}}\,.
        \label{condition}
    \end{equation}It is obvious from the above expression that for any arbitrary value of $z_h$ in between $z_h\rightarrow0$ and $z_h\rightarrow \infty$, the condition $\gamma\gg 1$ is not satisfied.
    \item If $\gamma=1$, i.e., $p_x= p_t$, then the equation (\ref{geodesic 2}) is not satisfied with $z_0=1$.
    \item Again, for $z_h=1$, $z_0=1$ is not a solution of the null geodesic equation for any arbitrary value of $\gamma$.
    \item The condition (\ref{condition}) clearly tells us that, for $z_0=1$ as a solution, $p_x$ cannot be less than $p_t$. Hence, with $p_x< p_t$, we must have $z_0\neq 1$.
    \item For any arbitrary value of $p_x$ and $p_t$; $z_0=0$ does not satisfy the geodesic equation.
\end{itemize}From the above conditions, it is evident that $z_0=1$ and $z_0=0$, i.e., the boundaries are not included as the consistent solutions of the null geodesic equation in our present study. For the relativistic asymptotically AdS black hole with $\xi=1$, the above procedure gives the null geodesic equation as \begin{equation}
    \dot{z}=z_0^2p_t\sqrt{1-(1-\frac{z_0^2}{z_h^2})\gamma^2}
\end{equation}For any arbitrary impact parameter $\gamma$, the turning point of this geodesic can be obtained only at $z_0=0$. Even for the relativistic case, the null geodesic equation is not satisfied if the corresponding turning point is at the horizon. So, for both the relativistic AdS and nonrelativistic Lifshitz black hole, the above method does not support the exact horizon of the geometry as the turning point of the null geodesic to carry out the expressions of the classical eikonal phase shift and Lyapunov exponent. 
 One may somewhat carry out a polynomial distribution by using a suitable numerical approach to see what values of $z_0$ the equation (\ref{geodesic 2}) can accommodate as its valid solution. We keep this computation for future extensions of our work. 
\subsection{Eikonal phase shift}
With the above analysis, here we will exhibit an explicit computation of eikonal phase shift as a nontrivial function of $z_0$ and $\xi$. With $p_t$ and $p_x$ as the conserved energy and momentum, respectively, for our system, the expression of the bulk phase shift is expressed as
\begin{equation}
    \delta\equiv p_t(\Delta t )-p_x(\Delta x)\,.
    \label{classical eikonal}
\end{equation}A fascinating point to observe is that the growth rate of the OTOCs found in the limit $u=0,~v=0,~t_2-t_1>>1$(i.e., the Regge limit for any double-sided Kruskal geometry) in section \ref{subsubsection } is $\frac{2\pi}{\beta}\left(t-t_*-\frac{\beta\mathcal{M}x}{2\pi}\right)$, where $\beta$, $t_*$ and $\mathcal{M}$ are nontrivial functions of $\xi$. For any fixed $\xi$, these quantities become constants. Let us assume that $t-t_*\sim\Delta t$ and the corresponding deflection occurs as $\Delta x\sim(x-0)$, then the OTOC growth rate represents the bulk phase shift for a 2-particle heavy-heavy-light-light scattering. With this scenario, we can infer that the conserved momenta $p_x$, energy $p_t$ and hence impact parameter $\gamma$ can be expressed nontrivially in terms of $\beta$, $t_*$ and $\mathcal{M}$, or in other words, in terms of $\xi$.
The deflections along the time and space directions can be found respectively from the expressions of $p_t$ and $p_x$ in (\ref{eom}) and the null geodesic equation (\ref{geodesic}) as
\begin{equation}
\begin{split}
    \Delta t\,&=\,2\int_{0}^{z_0}\frac{z^{2\xi-1}\dd z}{f(z)\sqrt{z^{2\xi}-\gamma^{2}z^{2}f(z)}}\,\\&=\,2a^{\frac{3}{2}}\int_{0}^{z_0} \frac{z^{2\xi-1}\dd z}{(a-z^{\xi+1})\sqrt{(az^{2\xi}-a\gamma^2z^2+\gamma^2z^{\xi+3})}}\,,
    \label{time deflection}
    \end{split}
\end{equation}
and
\begin{equation}
\begin{split}
    \Delta x\,&=\,2\int_{0}^{z_0}\frac{\gamma z\,\dd z}{\sqrt{z^{2\xi}-\gamma^{2}z^{2}f(z)}}\,\\&=\,2a^{\frac{1}{2}}\gamma\int_{0}^{z_0} \frac{z\,\dd z}{\sqrt{(az^{2\xi}-a\gamma^{2}z^{2}+\gamma^2z^{\xi+3})}}\,,
    \label{position deflection}
    \end{split}
\end{equation}where, $a=z_h^{\xi+1}$. Because of the intricate structure of the above integrands, we assume two limiting cases to derive them. Firstly, we take the near-boundary zone where $z\ll z_h$. Substituting (\ref{time deflection}) and (\ref{position deflection}) in the expression of the eikonal shift and simplifying, we get for the near boundary zone,
\begin{equation}
  \delta\sim
  \int_{0}^{z_0}\sqrt{z^{2\xi-2}-\gamma^2(1-\epsilon)} \dd z\approx  p_{t}\int_{0}^{z_0}\sqrt{z^{2\xi-2}-\gamma^2} \dd z\,,
\end{equation}where, $\left(\frac{z}{z_h}\right)^{\xi+1}=\epsilon\ll 1$ for $\xi\geq 1$.
Now defining a new integration variable $y=\frac{z}{z_0}$, the bulk phase shift is given by
  \begin{equation}
  \begin{split}
     \delta \sim &\int_{0}^{1} z_0\sqrt{z_0^{2\xi-2}y^{2\xi-2}-\gamma^2} \dd y \approx\\& i\gamma z_0 \, _2F_1\left(-\frac{1}{2},\frac{1}{2 \xi -2};\frac{2 \xi -1}{2 \xi -2};\frac{z_0^{2 \xi -2}}{\gamma ^2}\right)\,.
     \label{eikonal classical}
     \end{split}
  \end{equation}
  On the other hand, we take the near-horizon regime where $z\approx z_h$. In this regime, we can take two limiting cases for the impact parameter $\gamma$. When the energy $p_t$ is much greater than the momentum $p_x$, i.e.,  $\gamma <1$, the eikonal phase shift in near horizon case gives
 \begin{equation}
 \begin{split}
     \delta = &\frac{z_0^{\xi }}{\xi  (\xi +1)} \, _2F_1\left(1,\xi ;\xi +1;\frac{z_0}{z_h}\right) \\& =\frac{z_0^{\xi }}{\xi  (\xi +1)} \, _2F_1\left(1,\xi ;\xi +1;z_0\right)\,.
     \end{split}
  \end{equation}
 Again, for the small energy limit, i.e., $p_x\geq p_t$, the impact parameter  $\gamma \geq 1$. This, in the near-horizon zone, produces,
  \begin{equation}
  \begin{split}
   \delta& =  -\frac{4 \gamma  \left(\sqrt{z_0-z_h}-\sqrt{-z_h}\right) \sqrt{z_h} p_t}{\sqrt{\xi +1}}\\&= -\frac{4 \gamma  \left(\sqrt{z_0-1}- i \right)p_t}{\sqrt{\xi +1}}\,.
   \end{split}
  \end{equation}
  where we choose $z_h=1$ in the above expression for the sake of simplicity of our derivation. All the $\delta$'s derived in these limits inevitably happen to be nontrivial functions of the anisotropy index $\xi$. It is well-known in the backdrop of gravitational scattering that the real and imaginary nature of the eikonal phase shift respectively represent elastic and inelastic scatterings. Thus, our study of the classical eikonal phase clearly reveals that the nature of the gravitational scattering between the heavy-heavy-light-light particle collision in the Lifshitz black hole depends on the different regimes taken for the radial coordinate $z$ along with the choices of the impact parameter. For the near boundary case and the large impact parameter limit in the near horizon zone, the imaginary eikonal phase shift implies completely inelastic collisions, whereas the small impact parameter limit in the near horizon zone gives real values of $\delta $ that stand for elastic scattering. For elastic scattering in the near-horizon limit, the phase shift assumes a hypergeometric function, which is similar to the case with Regge behaviour of CFT. Due to the finite anisotropy of the chosen background, the hypergeometric function depends nontrivially on $\xi$. Hypergeometric function with nontrivial $\xi$-dependence has also been achieved for the inelastic scattering case in the near-boundary zone. This allows one to speculate about the dependencies of absorption cross-sections of different scattering scenarios on the finite anisotropy present in the black hole background. 
  \par Again, we use a WKB approximation to verify the consistency of the eikonal phase. Inserting the expressions of time and space deflections in \eqref{classical eikonal}, we can write, 
  \begin{equation}
      \delta=2 p_t \int_{0}^{z_0} \frac{\sqrt{z^{2\xi}-\gamma^2z^2f(z)}}{z f(z)}\, \dd z\,.
    \label{delta}
  \end{equation}One may check the emergence of the above eikonal phase via WKB approximation of the solution of the equation of motion of a scalar field. To do this, we consider a heavy scalar field $\phi$ propagating in the chosen Lifshitz background. In the background \eqref{Lifshitz BH 2}, the Klein-Gordon equation of the scalar can be written as
\begin{align}
   z^2f(z)\partial_{z}^2\phi+z^2\partial_{x}^2\phi+z^2f'(z)\partial_{x}^2\phi-\frac{z^{2\xi}}{f(z)}\partial_{t}^2\phi \nonumber\\-z^{\xi}f(z)\partial_{z}\phi+m^2z^{2+\xi}\phi=0\,.
\end{align}
Performing a Fourier transformation as
$$\phi=\Tilde{\phi}(z)\,e^{i(-p_t t +p_x x)}\,,$$
and taking an ansatz,
\begin{equation}\label{eik}
    \Tilde{\phi}(z)=e^{i p_t \chi (z)}\,,
\end{equation}
we can solve for $\chi(z)$. With the large energy ($p_t>>1$) limit, the EOM boils down to, 
\begin{equation}
    \left(\partial_{z}\chi(z)\right)^{2}=\frac{z^{2\xi}-\gamma^2 z^2 f(z)}{z^2f^2(z)}\,.
\end{equation}
Now, taking the positive term of $\partial_{z}\chi(z)$, we perform an integration and get a solution of $\chi(z)$ as
 \begin{equation}
   \chi(z)=\int_{0}^{z_0}\frac{\sqrt{z^{2\xi}-\gamma^2 z^2 f(z)}}{zf(z)}\,\dd z\,.
\end{equation}
where the terms of order $\mathcal{O}(p_{t}^{-1})$ can be ignored at large energy limit. This is indeed the leading term of the solution. We can see that the WKB approximation of the solution gives us the exact eikonal phase we recovered in \eqref{delta} at leading order in the large energy limit. 
\subsection{Lyapunov exponent}
The Lyapunov exponent ($\lambda_L$) is one of the key measures in understanding chaotic systems in classical phase space. It quantifies the rate at which the nearby trajectories diverge. Thus, it provides information on the system's sensitivity to the initial conditions. A positive Lyapunov exponent signifies the presence of chaos in a physical system.
Mathematically, it reads as
  $$\delta X(t)=\delta X(0)e^{\lambda_L t}\,,$$
  where $\delta X$ represents the separation between two nearby trajectories. In \cite{Cardoso:2008bp}, the authors established a connection between the Lyapunov exponent and the effective potential in the radial motion for both massive and massless particles. Building on this framework, we will calculate the Lyapunov exponent for a 2+1-dimensional planar Lifshitz black hole. Recently, \cite{Kumara:2024obd} conducted a similar analysis in the context of equatorial hyperscaling-violating black holes, treating Lifshitz black holes as a specific case. For a circular geodesic, the Lyapunov exponent is formulated in terms of the second derivative of the effective potential $V_{eff}$ for radial motion as \cite{Cardoso:2008bp,Kumara:2024obd}
\begin{equation}
    \lambda_L=\sqrt{-\frac{V_{eff}^{''}}{2\dot{t}^2}}\,.
    \label{lyapunov}
\end{equation}
For a massless particle, the geodesic equation is given in equation (\ref{geodesic}) from which we can calculate the effective potential  
\begin{equation}
    V_{eff}(z)=\dot{z}^2=(p_t)^2z^{2\xi+2}-(p_x)^2 z^4f(z)\,.
    \label{efctvpot}
\end{equation}
Where $t$ is the coordinate time. At the turning point $z=z_0$, the condition on circular orbit, i.e.,  $V_{eff}(z_0)=0$ and $V'_{eff}(z_0)=0$ yield
\begin{equation}
    \frac{p_t^2}{p_x^2}=z_0^{2-2\xi}f_0\,,
    \label{con1}
\end{equation}
and 
\begin{equation}
    z_{0}f_{0}'=2(\xi-1)f_{0}\,,
    \label{con2}
    \end{equation}
where $f_{0}=f(z_0)$. 
Upon substituting equations (\ref{con1}) and (\ref{con2}) in equation (\ref{efctvpot})
we get \begin{equation}
 V_{eff}''(z_0)=-z_0^2 p_x^2 \left(z_0^2 f_0''-2  f_0\left(2 \xi^2-5 \xi +3\right)\right)\,.
 \label{veffdp}
\end{equation}
Now, utilizing equations (\ref{eom}) and (\ref{veffdp}) in equation (\ref{lyapunov}), we derive the Lyapunov exponent for the null circular geodesic as
\begin{equation}
     \lambda_L=\sqrt{\frac{f_0\left(z_0^2 f_0''-2  f_0\left(2 \xi^2-5 \xi +3\right)\right)}{2z_0^{2\xi}}}\,.
     \label{lyap}
\end{equation}
Since an unstable circular geodesic meets the condition $V''_{eff}<0$, it follows from equations (\ref{veffdp}) and (\ref{lyap}) that $\lambda_L$  is real whenever the orbit is unstable. From equation \eqref{lyap}, it is clear that in AdS spacetime, i.e., for the anisotropic parameter $\xi = 1$, the Lyapunov exponent depends on both $z_0$ and $z_h$. For a horizon radius of $z_h=1$, the Lyapunov exponent indicates unstable orbits when $z_0 > 1$. In the near-horizon limit, the Taylor expansion of the function $f(z_0)$ around $z_0=z_h$ gives 
\begin{equation*}
\begin{split}
    f_0=f(z_0)=-\left[\frac{(1+\xi)}{z_h}(z_0-z_h)+\frac{\xi(\xi+1)}{2z_h^2}(z_0-z_h)^2\right. \\ \left.+\mathcal{O}[(z_0-z_h)^3]\right].
    \end{split}
\end{equation*}From this expression, we get 
$$f'(z_0)=-\left[\frac{(1+\xi)}{z_h}+\frac{\xi(\xi+1)}{2z_h^2}(z_0-z_h)+\mathcal{O}[(z_0-z_h)^2]\right],$$ and $$f''(z_0)=-\frac{\xi(\xi+1)}{2z_h^2}(z_0-z_h)+\mathcal{O}[(z_0-z_h)].$$ Substituting all these expressions in the equation (\ref{lyap}) and simplifying we get 
\begin{equation}
    \lambda_L=\frac{(1+\xi)}{z_0^{\xi}\sqrt{2}}\left[a(\xi)+b(\xi)\frac{z_0}{z_h}+\mathcal{O}\left(\frac{z_0^2}{z_h^2}\right)\right].
    \label{lyap 2}
\end{equation}where,\begin{align}
    a(\xi)=2\left(1-\frac{\xi}{2}\right)(1-\xi)(2\xi-3)\nonumber,\\b(\xi)=2(3-2\xi)(\xi-1)^2,\\c(\xi)=2\xi(\xi-1)(2\xi-3)-\xi\left(1-\frac{\xi}{2}\right)(1+\xi)\nonumber
\end{align} and other coefficients are also the functions of $\xi$. Now using the expression of the temperature of the Lifshitz black hole in (\ref{BH temperature}) we get from (\ref{lyap 2}),
\begin{equation}
\begin{split}
    \lambda_L=2\sqrt{2}\pi T\left[a(\xi)\left(\frac{z_h}{z_0}\right)^{\xi}+b(\xi)\left(\frac{z_h}{z_0}\right)^{\xi-1}\right. \\ \left.+\mathcal{O}\left[\left(\frac{z_h}{z_0}\right)^{\xi-2}\right]\right].
    \end{split}
\end{equation}
It is thus obvious that the classical Lyapunov exponent is also directly proportional to $T$, similar to the bound found for the Lyapunov exponent obtained in the three quantum methods discussed above for the asymptotically Lifshitz black hole. Therefore, to acquire the scenario of classical/quantum correspondence in our study, we get the relation
\begin{equation}
  a(\xi)\left(\frac{z_h}{z_0}\right)^{\xi}+b(\xi)\left(\frac{z_h}{z_0}\right)^{\xi-1}+\mathcal{O}\left[\left(\frac{z_h}{z_0}\right)^{\xi-2}\right]=\frac{1}{\sqrt{2}} ,
    \label{CM/QM}
\end{equation}which must be satisfied by a specific power series of $\left(\frac{z_h}{z_0}\right)^{\xi}$.

\section{Conclusion}\label{sec5}
We undergo a comparative analysis of three different holography-based approaches for the determination of quantum chaos. These include entanglement wedge reconstruction, out-of-time-ordered correlators and pole-skipping. We understand the plausible equivalence of these methods in the asymptotically Lifshitz black hole with arbitrary anisotropy index. 
In the entanglement wedge method, we construct an extremal hypersurface as a constant time slice, famously called the Ryu-Takayanagi surface, that extracts out the butterfly velocity at late time. The butterfly velocity exhibits monotonically increasing behaviour with both the anisotropy index and the finite temperature of the chosen background. For the unit anisotropy index, it is expected to reduce to that for the planar AdS black hole. For the derivation of OTOCs, we explicitly construct the double-sided Kruskal geometry corresponding to the asymptotically Lifshitz black hole. 
We introduce a gravitational shockwave backreacting in the Kruskal version of our background. The functional form of the shockwave eventually attains exponential structure and gives rise to the OTOCs. It further shows that the system is chaotic and the rate of chaotic growth is specified by the Lyapunov exponent. This is somewhat consistent with the previously obtained scenario for any generic perturbed holographic two-sided gravity background. The butterfly velocity and Lyapunov exponent obtained from the shockwave profile are exactly similar to those from the entanglement wedge method. Finally, in the method of pole-skipping, we again observe the exact same dependency of butterfly velocity as well as the Lyapunov exponent on the anisotropy. As a consequence of the exact matching of these salient chaotic features, we claim that all these chosen methods are equivalent and definitive to understanding quantum chaos for the class of asymptotically Lifshitz black holes. Various chaos related studies have been done in anisotropic backgrounds. In \cite{Sircar:2016old}, the authors have studied the mutual information in asymptotically Lifshitz black holes. They have explicitly studied how mutual information behaves in the presence of a shock wave. As introduced in our subsection \eqref{otoc_subsec}, the shift $\alpha$ and mutual information relation have been explored. In \cite{Giataganas:2017koz}, the authors have studied the transport and diffusion properties in anisotropic theories. They have shown that, for anisotropic theories, the longitudinal component of butterfly velocity diffuses slower, whereas the transverse component propagates faster. In addition, in \cite{Jahnke:2017iwi}, the author has studied the chaos-related properties of anisotropic systems in the presence of a shock wave. In this case, the transverse component of the entanglement velocity decreases with anisotropy, while the parallel component of the velocity increases with anisotropy. Their results show that anisotropy increases the two-sided entanglement.
However, the butterfly velocity we found is $v_B\sim T^{1-\frac{1}{\xi}}$ \eqref{butterfly_temp} which matches with \cite{Blake:2016wvh, Roberts:2016wdl}. In our work, we took a step further and computed the chaotic parameters in the Lifshitz background (which is anisotropic) by three methods. And we found the equivalence among these methods. 
 \par We conduct a further investigation of the eikonal phase shift due to heavy-heavy-light-light gravitational scattering and compute the Lyapunov exponent at the classical level by using the bulk metric both in near-boundary and near-horizon regimes. A detailed study of the associated null geodesics serves this purpose. We infer a compatible connection of the quantum OTOCs in the leading order of the series expansion of its exponential form with the classical eikonal phase shift in the bulk. This reveals that the leading order term of OTOCs at the classical level can be studied as nontrivial functions of the turning point of the null geodesic as well as the anisotropy index. Interestingly, due to the presence of arbitrary anisotropy in our case, the eikonal phase becomes real or imaginary depending on the impact parameter and the near-horizon and near-boundary limits of the black hole. This may allow one to explore the absorption cross-sections for both elastic and inelastic scatterings in terms of the arbitrary anisotropy index of the asymptotically Lifshitz black hole. We successfully verified the eikonal phase by using the WKB approximation of the solution of the Klein-Gordon equation, at least up to the leading order. The classical Lyapunov exponent also appears to be a nontrivial function of the anisotropy index $\xi$. It is observed that, in the near-horizon limit, there occurs an emergent classical/quantum correspondence for the class of Lifshitz theories for a specific convergent power series of $\left(\frac{z_h}{z_0}\right)^{\xi}$ with $\xi$-dependent coefficients. The series must converge at $\frac{1}{\sqrt{2}}$ to achieve an exact matching between classical and quantum Lyapunov exponents. In addition to these, we further remark on some obvious restrictions on the possible extrema of the null geodesics, where we obtain inconsistent natures of the classical eikonal phase shift and Lyapunov exponent. It would be further interesting to explore the possible connection between the phase shift and the Lyapunov exponent, which may eventually lead to classical/quantum correspondence in the context of chaos in
Lifshitz black hole. As another potential future research, it would be fascinating to have a more elaborate investigation of the classical/quantum correspondence in the context of a generic class of holographic Lifshitz theories.  One can also go for an explicit evaluation of the OTOCs from the systematic construction of the dual thermofield double states of the finite temperature Lifshitz field theory and subsequently study different chaotic features on the field theory side. We observed that our specific methods hold the robustness of growing chaos for increasing anisotropy in the Lifshitz black hole geometry. Nevertheless, one can intuitively think of a system where the symmetry breaking suppresses the chaos considerably. Such understanding still lacks explicit analysis in the non-Lorentzian framework like Lifshitz geometries. It would be intriguing to explore if such spacetime exists as a solution of the Lifshitz class of gravity theories. Furthermore, there are seminal works\cite{Lee:2010qs, Lee:2010ii, Lee:2011zzf}, where the class of anisotropic Lifshitz gravity on emergent curved backgrounds are conjectured to be a reasonable dual for strange metals. In this context, one can expect from our study a similar equivalence of the prescribed holographic methods for quantum chaos in the holographic strange metal models and also an emergence of classical/quantum correspondence. Another illuminating idea is to understand chaos in a holographic sense for the Lifshitz field theories living on curved geometries. Previously, such theories, including self-interaction, were renormalizable for $ \xi=3$ \cite{LopezNacir:2011mt}. One can infer a probable holographic dual gravity background which follows Lifshitz scale symmetry and contains the boundary as a curved geometry, such as a sphere. It will be rather fascinating to undergo the analysis of quantum and classical chaos and understand how these properties may be affected due to such bulk geometries. We hope to come back with some of these ideas in the near future.
\acknowledgments
We would like to thank Debaprasad Maity for his valuable suggestions regarding this work. A special thanks to Diandian Wang and Joydeep Chakravarty for some helpful suggestions in our latest draft. BB is grateful to Arnab Kundu for hosting a visit to SINP, Kolkata, and to Pankaj Chaturvedi for facilitating a visit to NIT, Silchar, where part of this work was conducted. Additionally, BB would like to thank Akhil Sivakumar for various insightful suggestions on this work. AC appreciates Bibhas Ranjan Majhi for hosting a visit to IIT Guwahati, which marked the initiation of this work. AC would like to thank the  Research project supported by the program "Excellence initiative – research university" for the AGH University for providing funds to carry out the above work. AC also acknowledges the financial support from the NSTC, Taiwan (R.O.C.), under grant number 110-2112-M007-015-MY3, which funded a significant portion of this research. Finally, AC extends sincere thanks to Chong-Sun Chu, Dimitrios Giataganas, Himansu Parihar, and Jaydeep Kumar Basak for their insightful discussions on various aspects of Lifshitz theories, which were extremely beneficial to this work.
\twocolumngrid
\bibliographystyle{apsrev4-1}
\bibliography{prd_lifshitz}

\end{document}